\documentclass[conference]{IEEEtran}

\usepackage[utf8]{inputenc}
\usepackage{amsmath,amssymb}
\usepackage{array}
\usepackage{booktabs}
\usepackage{graphicx}
\usepackage{multirow}
\graphicspath{{figures/}}
\usepackage{hyperref}
\usepackage{listings}
\usepackage{url}
\usepackage{tikz}
\usepackage{xcolor}
\usepackage{soul}
\definecolor{myyellow}{RGB}{255, 255, 150}
\sethlcolor{myyellow}

\hypersetup{hidelinks}

\definecolor{codebg}{RGB}{248, 248, 248}
\definecolor{codeframe}{RGB}{220, 220, 220}
\lstdefinestyle{capsealcode}{
  basicstyle=\ttfamily\small,
  breaklines=true,
  frame=single,
  framerule=0.5pt,
  rulecolor=\color{codeframe},
  framesep=2mm,
  backgroundcolor=\color{codebg},
  tabsize=2,
  columns=fullflexible,
  keepspaces=true,
  showstringspaces=false
}
\title{CapSeal: Capability-Sealed Secret Mediation for Secure Agent Execution}

\author{
    \IEEEauthorblockN{Shutong Jin\textsuperscript{1}, Ruiyi Guo\textsuperscript{2}, and Ray C. C. Cheung\textsuperscript{1}}
    \IEEEauthorblockA{\textsuperscript{1}City University of Hong Kong, Hong Kong \\
    \textsuperscript{2}Beijing Foreign Studies University} 
}
\begin{document}

\maketitle

\begin{abstract}
Modern AI agents routinely depend on secrets such as API keys and SSH credentials, yet the dominant deployment model still exposes those secrets directly to the agent process through environment variables, local files, or forwarding sockets. This design fails against prompt injection, tool misuse, and model-controlled exfiltration because the agent can both use and reveal the same bearer credential. We present \textbf{CapSeal}, a capability-sealed secret mediation architecture that replaces direct secret access with constrained invocations through a local trusted broker. CapSeal combines capability issuance, schema-constrained HTTP execution, broker-executed SSH actions, anti-replay session binding, policy evaluation, and tamper-evident audit trails. We describe a Rust prototype integrated with an MCP-facing adapter, formulate conditional security goals for non-disclosure, constrained use, replay resistance, and auditability, and define an evaluation plan spanning prompt injection, tool misuse, and SSH abuse. The resulting system reframes secret handling for agentic systems from ``hand the model a key'' to ``grant the model a narrowly scoped, non-exportable action capability.''
\end{abstract}

\section{Introduction}

The integration of autonomous AI agents has catalyzed a paradigm shift in software development and production, offering unprecedented gains in efficiency and task automation. However, this rapid adoption has outpaced the evolution of security frameworks. Traditional defensive measures—such as static encryption and digital signatures designed for human-mediated workflows—are increasingly inadequate in an era of agentic autonomy. As agents transition from passive assistants to active decision-makers with access to sensitive environments, the attack surface has expanded in both complexity and scale.

The current landscape is fraught with emerging threats specifically targeting agentic logic and integrations. Notable among these are:
\begin{itemize}
    \item \textbf{Prompt Injection:} Exploiting Large Language Model (LLM) vulnerabilities to hijack agent intent \cite{Greshake_2023,Lee2024promptinfection}.
    \item \textbf{Skill Poisoning:} Corrupting the functional capabilities or ``skills'' an agent retrieves to perform tasks \cite{Chen_2024agentpoison,Wang_2024badagent}.
    \item \textbf{MCP (Model Context Protocol) Poisoning:} Injecting malicious context or instructions through standardized communication protocols \cite{Hou_2026mcp,Wang_2026mcptox}.
    \item \textbf{Supply Chain Attacks:} Compromising the external libraries and toolkits that agents rely upon \cite{Williams_2025supplychain,Przymus_2025xzutils}.
\end{itemize}

We posit that the risks inherent in the agent era are significantly more severe than those of the preceding pre-agent era for three primary reasons. First, agents rely heavily on dynamic, online toolkits and third-party sources, increasing exposure to external vulnerabilities. Second, the democratization of cyberattacks means that adversaries are now utilizing coding agents themselves to automate vulnerability discovery and exploit generation, making the execution of attacks far more accessible. Finally, the growing reliance on LLM-based code review cannot be fully trusted, as these models often fail to detect sophisticated, multi-step adversarial logic. Recent real-world evidence, such as the supply chain compromises in \texttt{axios} and \texttt{liteLLM}, alongside skill poisoning instances in \texttt{Clawhub}, demonstrate that attackers are advancing at the same velocity as the developers building these agentic systems.

The core motivation of this research is that agent-targeting attacks cannot be effectively mitigated through traditional means, as the attack-and-defense warfare is growing rapidly alongside AI capabilities. We therefore approach this problem from a \emph{cryptographical} perspective: can we protect an agent's most valuable data by default?

While standard \emph{at-rest} encryption provides a baseline of safety, it is insufficient for autonomous systems. In an agentic workflow, the agent must frequently utilize these secrets to authorize transactions, generate signatures, or access functional APIs. Consequently, the moment of greatest vulnerability is during \emph{use-time}. 

To bridge this gap, this paper presents \textbf{CapSeal}, a capability-sealed secret mediation architecture that addresses this use-time exposure by replacing direct credential access with constrained, broker-mediated invocations. The agent never obtains the secret itself; instead, it requests a session-bound handle for a specific, policy-evaluated intent, and a local trusted broker mediates all credential-bearing actions through typed execution paths. The broker enforces schema constraints, redacts outputs, tracks anti-replay state, and records decisions in a tamper-evident audit chain~\cite{crosby-wallach}.

\textbf{This paper makes four contributions.}
\begin{itemize}
    \item We articulate the \emph{use-time secret exposure} problem for agent systems and argue that direct secret mounting is incompatible with prompt-robust agent execution.
    \item We present \textbf{CapSeal}, a capability-sealed mediation architecture for session establishment, capability issuance, invocation, revocation, and audit proof export.
    \item We design two concrete capability realizations: schema-confined HTTP execution with credential injection and broker-executed SSH commands with host and command-template constraints.
    \item We define a reproducible evaluation methodology across benign tasks, prompt injection, tool misuse, and SSH abuse, with MCP tool poisoning isolated as an explicit extension experiment.
\end{itemize}
In addition, we provide a same-harness latency comparison against direct execution and two external mediation baselines, allowing CapSeal's runtime cost to be assessed under identical HTTP and SSH tasks.

\section{Background and Threat Model}

\subsection{Why direct secrets fail in agent systems}

Bearer credentials are unsafe when handed to a component that is both semantically steerable and externally connected. RFC 6750 explicitly warns that a bearer token grants access to any party that possesses it \cite{rfc6750}. In agent systems, possession is not limited to memory reads: a prompted model can transform, summarize, paraphrase, or exfiltrate the same credential through tool parameters, logs, or follow-on instructions \cite{li2024personal}.

This risk is amplified by the agent tool plane. MCP and similar frameworks standardize tool discovery and invocation, which improves developer ergonomics but also expands the attack surface to tool descriptions, tool-choice prompts, and output handling \cite{mcp-spec}. Consequently, the system must defend not only against a malicious final tool invocation, but also against earlier steering that changes what tool the model chooses to call \cite{prompt-injection-tool-selection,ipiguard2025}.

\subsection{Adversary and trust assumptions}

CapSeal adopts a deliberately conditional threat model.

\textbf{Adversary capabilities.}
\begin{itemize}
    \item The adversary can influence agent prompts, retrieved context, or tool descriptions \cite{taskshield2024,melon2025}.
    \item The adversary can cause the agent to request capabilities or submit crafted invocation payloads.
    \item The adversary may control remote services or network paths outside the local broker boundary.
\end{itemize}

\textbf{Trusted computing base.}
\begin{itemize}
    \item The local broker, policy engine, secret store, and audit subsystem form the minimal TCB \cite{vault-transit,aws-kms}.
    \item The operating system correctly enforces local process isolation and Unix-domain-socket peer identity.
    \item The adversary does not have local root or kernel compromise in the v1 model.
\end{itemize}

\textbf{Out of scope.}
\begin{itemize}
    \item Memory scraping or binary replacement by a local root adversary.
    \item Side channels below the abstraction level of broker-visible protocol events.
    \item Cross-host broker federation, multi-tenant isolation, and hardware-bound attestation.
\end{itemize}

\section{CapSeal Architecture and Protocol}
\label{sec:architecture}

The design of CapSeal is governed by four primary security objectives, which define the trust boundaries between the agent, the broker, and external services.

\textbf{G1: Secret Non-disclosure.} Under our established trust assumptions, the agent must never obtain the secret plaintext or any functional equivalent (e.g., replayable session tokens or raw private keys). The architecture ensures that secrets are only materialized within the broker's isolated execution context.

\textbf{G2: Fine-grained Policy Enforcement.} Every invocation must satisfy a multi-dimensional constraint check. The broker enforces least-privilege access by validating requests against issued capability specifications, including destination host/path restrictions, command templates, rate-limiting quotas, and mandatory step-up authentication.

\textbf{G3: Temporal and Contextual Binding.} To prevent credential hijacking, invocation messages must be resistant to replay attacks and context mis-binding. Capabilities are bound to specific sessions and channels, utilizing anti-replay state to ensure that captured traffic cannot be reused across different temporal or logical contexts.

\textbf{G4: Tamper-Evident Accountability.} Every security-relevant event—including issuance, invocation, and revocation—is recorded in an append-only, integrity-protected structure. This allows for asynchronous consistency checking and provides a verifiable audit trail \cite{rfc9162,crosby-wallach}.

Critically, CapSeal's guarantees are stronger than simple "at-rest" encryption but narrower than a full hardware-root-of-trust (absent TEE extensions). We assume a boundary where an agent, while potentially malicious or compromised, is logically isolated such that it cannot exfiltrate secrets and can only trigger actions explicitly authorized by the broker.

\begin{figure}[t]
\centering
\includegraphics[width=0.48\textwidth]{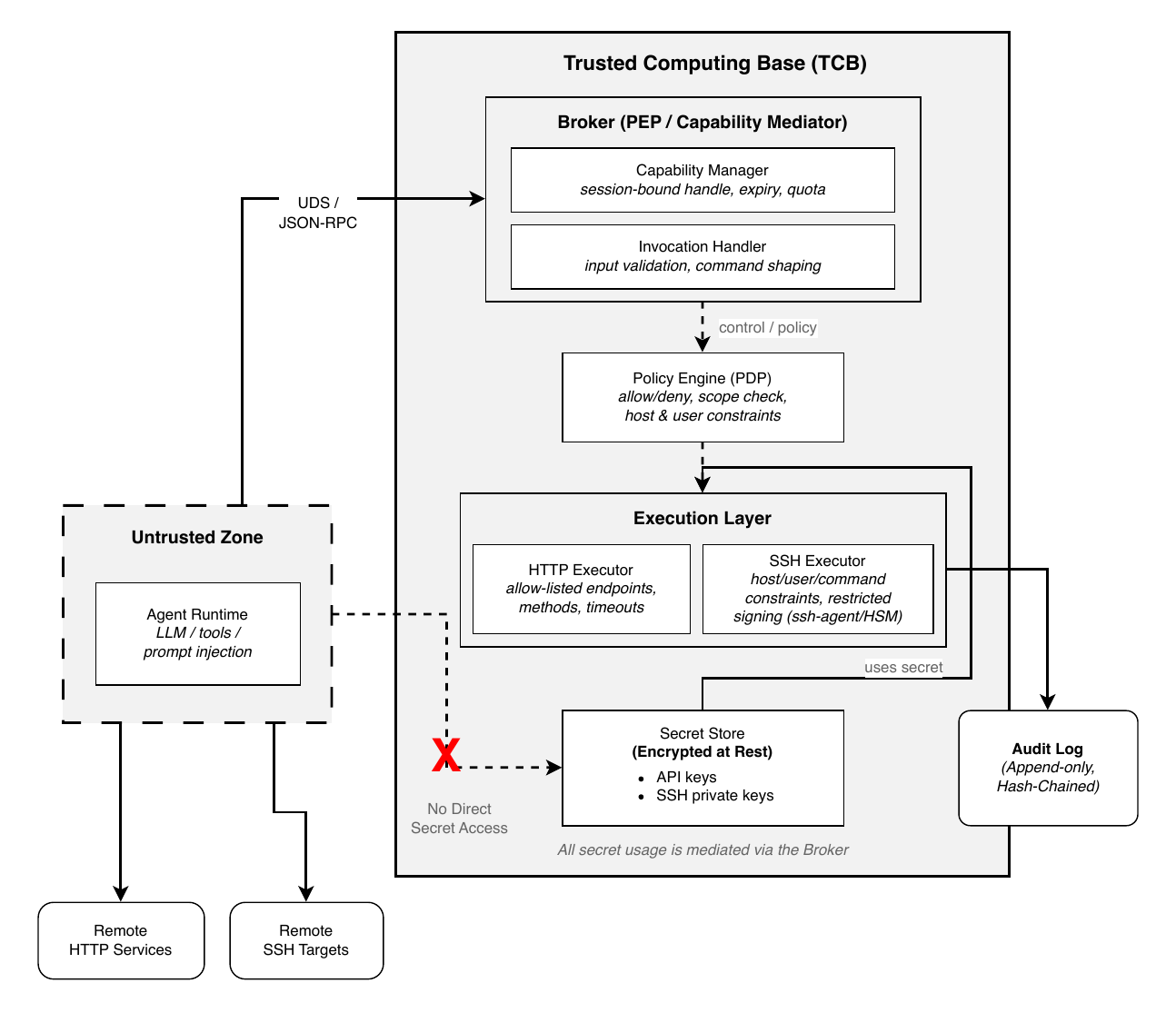}
\caption{CapSeal architecture and trust boundary. The broker acts as a reference monitor; agents submit intents and payloads, while the broker manages secret injection and execution.}
\label{fig:architecture}
\end{figure}

As illustrated in Figure~\ref{fig:architecture}, the CapSeal architecture centers on a fundamental decoupling: the agent expresses \textit{intent}, but only the broker performs \textit{execution}. This separation transforms the broker into a formal Policy Enforcement Point (PEP) and Policy Decision Point (PDP). The agent is restricted to requesting capabilities and submitting payloads for redaction or processing, while the secret material remains strictly confined to the broker-side execution path.

\subsection{Architectural Components}

The CapSeal control plane is structured as a mediated pipeline. Rather than providing a flat set of tools, the system enforces a strict state machine: a client must first establish a secure session, obtain a narrowly-scoped capability handle through policy approval, and only then proceed to invocation. This architecture ensures that protocol messages serve as sequential security gates rather than simple API endpoints.

\subsection{Request Lifecycle and Protocol Flow}

The execution lifecycle begins with the agent interacting with an MCP-compatible runtime. To the agent, the system appears as a set of standard tools; however, these tools are mediated by a CapSeal adapter. As shown in Table~\ref{tab:rpc_mapping}, the current prototype maps high-level MCP verbs to low-level broker operations implemented in Rust. The adapter facilitates communication via JSON-RPC over Unix Domain Sockets (UDS), maintaining a clean separation between the agent's high-level tool use and the broker's rigorous session management.

The CapSeal protocol processes each request through six distinct stages:
\begin{enumerate}
    \item \textbf{Registration:} The \texttt{register} operation establishes a session context, binding subsequent actions to a verified transport identity.
    \item \textbf{Capability Request:} The \texttt{req\_cap} operation declares an intent, specifying the required capability type and associated scope constraints (e.g., allowed hosts or paths).
    \item \textbf{Policy Evaluation:} The PDP evaluates the request against active security policies, determining if the authority should be granted, denied, or require multi-factor ``step-up'' approval.
    \item \textbf{Invocation:} The agent uses the issued handle via \texttt{invoke}, providing necessary payloads and anti-replay metadata.
    \item \textbf{Mediated Execution:} The broker validates the invocation against session state and capability constraints, injects the required secrets, and executes the action on the agent's behalf.
    \item \textbf{Audit Export:} Finally, \texttt{audit.prove} generates cryptographic evidence of the transaction for the append-only ledger.
\end{enumerate}

\begin{table}[t]
\caption{Protocol operations, security purpose, and primary enforcement checks.}
\label{tab:rpc_mapping}
\centering
\small
\begin{tabular}{p{0.20\linewidth} p{0.30\linewidth} p{0.41\linewidth}}
\toprule
\textbf{Operation} & \textbf{Security Purpose} & \textbf{Primary Enforcement Checks} \\
\midrule
\texttt{register} & Session Establishment & Transport binding, peer-cred verification \\
\texttt{req\_cap} & Authority Granting & Policy match, intent parsing, step-up requirements \\
\texttt{invoke} & Mediated Execution & Session validity, anti-replay, TTL, quota, scope validation \\
\texttt{revoke} & Authority Rescission & State update, immediate handle invalidation \\
\texttt{audit.prove} & Accountability & Consistency proof generation, chain verification \\
\bottomrule
\end{tabular}
\end{table}

The \texttt{scope} field within a capability is the primary mechanism for authority narrowing. By restricting parameters—such as pinning an HTTP request to a specific POST method on \texttt{api.example.com} and validating the payload against a JSON Type Definition schema \cite{rfc8927}—the broker ensures the agent cannot deviate from the pre-authorized intent. If an agent attempts to manipulate the destination or the payload structure, the broker terminates the request before the secret is ever exposed to the network stack.

\subsection{Session Binding and Replay Resistance}

To satisfy goals \textbf{G2} and \textbf{G3}, CapSeal implements strict session binding. The broker utilizes UDS peer-credential extraction to identify the calling process, preventing identity self-assertion. A capability handle is cryptographically useless outside the specific channel and session for which it was generated. 

Replay protection is enforced through an \texttt{AntiReplay} structure containing a monotonically increasing sequence number, a unique nonce, and a millisecond-precision timestamp. An invocation is only accepted if it satisfies four concurrent conditions: (i) it is bound to an active session, (ii) it references a valid, non-expired handle, (iii) it passes freshness and nonce-tracking checks, and (iv) it falls within the allotted call quota. This multi-layered validation ensures that authority remains a transient, context-bound privilege rather than a reusable credential.
\begin{figure*}[t]
\centering
\includegraphics[width=0.8\textwidth]{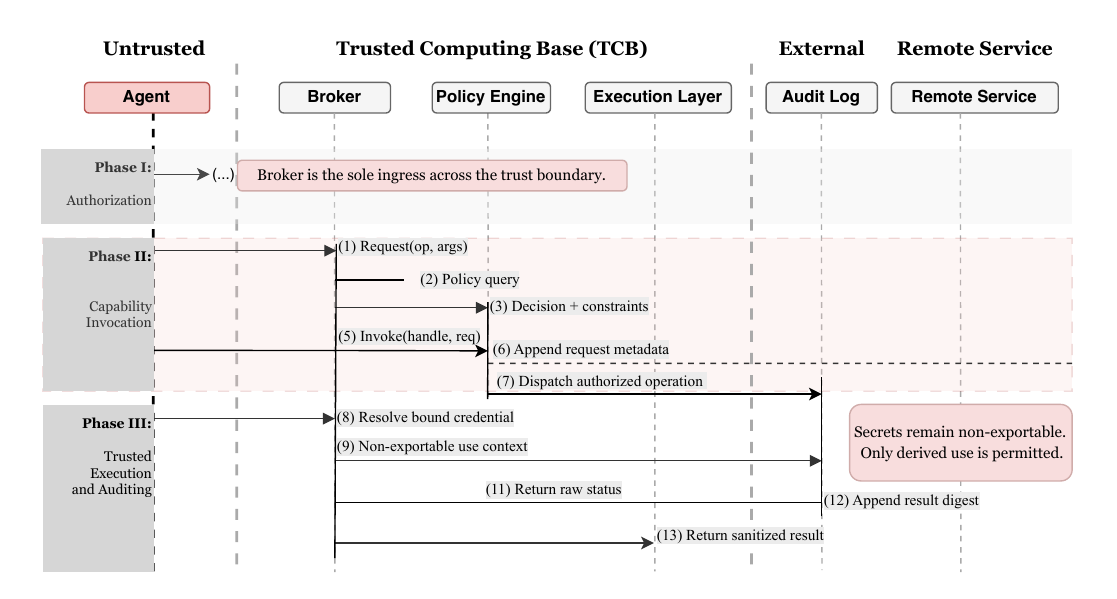}
\caption{CapSeal request lifecycle from capability request to broker-mediated execution and audit proof export.}
\label{fig:protocol}
\end{figure*}

\section{Capability Realizations}

\subsection{HTTP capability}

The HTTP capability is designed to behave as a constrained request constructor rather than a general-purpose forward proxy. That distinction matters because a generic proxy would still hand the agent broad routing power, making it easy to redirect requests, reshape payloads, or smuggle credentials through loosely checked parameters. CapSeal instead treats an HTTP capability as pre-authorized authority over a narrow method/host/path surface whose request body can itself be semantically confined.

The following example illustrates how an agent requests a capability for calling an OpenAI-like API. The request declares the agent's intent (\texttt{http\_call\_openai\_like}), specifies the credential resource (\texttt{OPENAI\_API\_KEY}), and defines narrow scope constraints that the broker will enforce during invocation.

\begin{figure}[htbp]
\centering
\caption{Capability request for schema-constrained HTTP action.}
\label{lst:reqcap}
\begin{lstlisting}
{
  "jsonrpc": "2.0",
  "id": "2",
  "method": "capseal.req_cap",
  "params": {
    "session_id":       "sess_4d7f",
    "intent":           "http_call_openai_like",
    "cap_type":         "HTTP_PROXY",
    "resource": {
      "secret_id":      "OPENAI_API_KEY"
    },
    "scope": {
      "method":         ["POST"],
      "host":           "api.example.com",
      "path_template":  "/v1/chat/completions",
      "body_schema_ref": "jtd:ChatCompletionRequest.v1"
    }
  }
}
\end{lstlisting}
\end{figure}

The narrowing path is layered. The broker first validates method, host, and path against the issued scope, then enforces payload-byte limits and header allowlists, rejects sensitive caller-supplied authorization headers, and injects the credential internally. Only after those network-level checks pass does it validate the body against the declared schema reference. Schema validation operates on typed request definitions, ensuring that the request body conforms to the permitted structure. This is the point where CapSeal moves beyond simple network mediation into semantic mediation: the agent is not merely restricted to a destination, but to a permitted request shape.

The HTTP executor enforces these constraints through progressive validation. The system first validates network-level restrictions (method, host, path), then enforces payload size limits and header policies, sanitizes credentials in responses, and injects authentication tokens internally. Schema validation failures result in denial before any credential use, ensuring fail-closed behavior. The broker mediates communication with external endpoints, treating unreachable or unauthorized destinations as constraint violations that trigger denial. When audit evidence generation is enabled, the broker records request-side artifacts—request structure, policy decisions, and response metadata—without exposing the underlying credential to the agent. This ensures that even if policy permits a capability, the executor prevents misuse through type-specific validation.

\subsection{SSH capability}

SSH uses a stricter realization because the threat surface is stronger. If CapSeal merely forwarded an SSH agent or exposed a reusable signing interface, the agent would still hold a powerful ambient channel whose misuse is hard to bound. For that reason, CapSeal adopts a broker-exec model: the broker executes a narrowly constrained remote action, and the agent never directly holds SSH secret material or a forwarding-capable socket \cite{openssh-config,openssh-add}.

The following request is the exact SSH capability issuance shape used by the MCP adapter in our real Docker end-to-end experiment trace.

\begin{figure}[htbp]
\centering
\caption{SSH capability request from real MCP end-to-end experiment transcript.}
\label{fig:ssh_reqcap}
\begin{lstlisting}
{
  "jsonrpc": "2.0",
  "id": 3,
  "method": "tools/call",
  "params": {
    "name": "capseal.req_cap",
    "arguments": {
      "intent": "ssh_mcp_e2e",
      "cap_type": "SshExec",
      "resource": {
        "secret_id": "SSH_PROD_KEY"
      },
      "scope": {
        "Ssh": {
          "host": "sshd",
          "user": "capseal",
          "command_prefix": [
            "ssh",
            "-i"
          ],
          "max_arguments": 3,
          "known_hosts_pin": "ssh-ed25519"
          "max_output_bytes": 2048
        }
      },
      "step_up": "None"
    }
  }
}
\end{lstlisting}
\end{figure}

The enforcement path again narrows authority step by step. The capability first constrains the remote host and user, then restricts execution to an approved command-prefix template with bounded arguments, forbids forwarding, and caps output size. The culminating control is host authenticity: the executor compares the presented host key against the capability's \texttt{known\_hosts\_pin}. That final comparison is what prevents a session from remaining valid if the agent is redirected to a different SSH endpoint that happens to fit the same superficial hostname or command shape.

The SSH executor enforces these constraints through a progressive validation sequence. The system verifies host and user allowlists, validates command invocations against approved templates, limits argument counts, prohibits agent forwarding, and bounds output sizes. Critically, the executor performs cryptographic host key verification, comparing the presented host identity against the pinned key specified in the capability. This cryptographic binding prevents capability misuse even if an adversary controls a host matching the requested hostname or command pattern. The architecture deliberately adopts broker-exec over signing-oracle delegation because the latter's security depends on forwarding constraints that are difficult to enforce statefully and provide weaker isolation than direct broker mediation.

\subsection{Invocation Control Flow}

Regardless of capability type, every invocation follows a defense-in-depth validation sequence before the broker allows secret-bearing execution. The following logic illustrates the ordered checks that enforce session binding, anti-replay guarantees, quota limits, policy decisions, and type-specific executor constraints.
This validation sequence realizes a defense-in-depth architecture where multiple independent security layers collectively ensure safe credential use. Session and capability lookups provide contextual validity. Replay detection through monotonic sequence numbers, unique nonces, and timestamp bounds prevents reuse attacks. Revocation, expiry, and quota checks enforce temporal and usage constraints. Policy evaluation enables runtime authorization decisions that can adapt to threat context. Executor constraint validation enforces the type-specific narrowing rules described above—HTTP schema validation or SSH command-template enforcement. Only when all checks pass does the broker inject credentials and perform the action, followed by an immutable audit log entry \cite{crosby-wallach}. This layered design ensures that compromising a single validation layer does not grant the adversary access to credentials or unrestricted capability use—each layer provides an independent security barrier that must be satisfied before secret-bearing execution occurs.

\begin{table*}[t]
\caption{Authority narrowing mechanisms in CapSeal.}
\label{tab:constraints}
\centering
\begin{tabular}{p{0.18\linewidth} p{0.20\linewidth} p{0.22\linewidth} p{0.30\linewidth}}
\toprule
Capability area & Authority source & Enforcement point & Enforcement mechanisms \\
\midrule
Transport/session & Session-bound capability handle & UDS transport and broker session state & Peer credential binding, anti-replay validation (monotonic sequence, nonce uniqueness, timestamp bounds), TTL/quota/revocation checks \\
HTTP & Broker-held API credential & HTTP executor & Network-level restrictions (method, host, path), payload size limits, header policies, credential injection, schema validation, audit evidence generation \\
SSH & Broker-held SSH material & SSH executor & Host/user allowlists, command template validation, argument limits, agent forwarding prohibition, output bounds, cryptographic host key verification \\
Policy & Contextual authorization decision & Policy engine & Runtime policy evaluation, allow/deny/step-up decisions, auditable policy trace \\
Audit & Append-only evidence chain & Audit and secret subsystems & Cryptographic audit chain, optional cryptographic signing, encrypted secret storage \\
\bottomrule
\end{tabular}
\end{table*}

\section{Integration and Extension}
\label{sec:integration}

To demonstrate the versatility and robustness of the CapSeal framework, we explore its integration with emerging industry standards, hardware-level security primitives, and diverse deployment scenarios. This section details how CapSeal extends its capability-based mediation to the Model Context Protocol (MCP), Trusted Execution Environments (TEE), and Internet of Things (IoT) ecosystems.

\subsection{Standardizing Agent-Tool Interaction via MCP}

The \textit{Model Context Protocol (MCP)} has emerged as a standard boundary for decoupling agent runtimes from external toolsets. CapSeal leverages this protocol to formalize the interface between the agent's intent and the broker's enforcement. By adopting MCP, CapSeal effectively separates the \textit{request for action} from the \textit{secret-bearing execution}, ensuring that agents interact with a familiar, high-level protocol surface rather than handling low-level, sensitive credentials.

This integration enables CapSeal to be seamlessly plugged into mainstream agent toolchains. Instead of receiving raw, reusable API keys, the agent invokes approved actions through a standardized MCP layer. This layer acts as a specialized adapter that translates protocol-compliant requests into internal CapSeal capability operations. Crucially, while the MCP layer facilitates interoperability at the edge, the CapSeal broker maintains its role as the ultimate policy and execution authority, keeping security-critical decisions isolated from the agent's immediate environment.

\subsection{Hardening the Broker with Trusted Execution Environments (TEE)}

While CapSeal provides logical isolation, the broker itself remains a high-value target. To mitigate the risk of host-level software compromise, we extend CapSeal with support for \textit{Trusted Execution Environments (TEEs)}. By utilizing hardware-enforced isolation and remote attestation, we move sensitive security logic---such as key management and policy evaluation---into a protected enclave.

Integrating TEEs into the CapSeal architecture significantly raises the assurance level of the system. Specifically, the broker's secret-handling routines and session-control logic are executed within an attested TEE boundary, shielding them from potentially compromised operating systems or hypervisors. This ``hardware-root-of-trust'' approach reinforces the integrity of the broker without necessitating changes to the existing capability protocol or the developer workflow. From the agent's perspective, the transition to a TEE-backed broker is transparent, yet the underlying protection against lateral movement and memory introspection is substantially strengthened.

\subsection{Extending Mediation to IoT and Edge Computing}

The challenges of \textit{Internet of Things (IoT)} and edge deployments---characterized by heterogeneous hardware, intermittent connectivity, and the need for device-scoped authority---align closely with CapSeal’s design philosophy. In these settings, the risk of physical device compromise necessitates a transition away from long-lived, broad-scope credentials toward narrow, auditable control over secret use.

CapSeal adapts to IoT environments by functioning as a distributed mediation layer. In this model, each edge node or IoT device is granted highly constrained capabilities rather than direct access to central secrets. This architecture minimizes the \textit{blast radius} of a single node compromise; an attacker gaining control of a device is limited to the specific, pre-authorized actions defined by its active capabilities. Furthermore, CapSeal's centralized audit and revocation mechanisms allow administrators to manage fleet behavior in real-time, providing a scalable framework for orchestrating complex, secure operations in distributed physical settings.
\section{Prototype Implementation}

We realize the capability-mediated architecture as a broker-centered system with integrated evaluation infrastructure. The implementation demonstrates the full security lifecycle: session-aware capability issuance, typed executor enforcement, policy-mediated approval, secret confinement within the broker boundary, and auditable evidence export. This end-to-end realization is essential because the security guarantees depend on the coordinated interaction among these components rather than on any individual mechanism in isolation.

\subsection{Prototype Overview}

The architecture centers on the broker component, which manages the complete capability lifecycle: session registration, capability request and issuance, invocation, revocation, and audit proof generation. Supporting subsystems include session management, typed executors for HTTP and SSH capabilities, the policy evaluation layer, secret storage backends, audit chain maintenance, and the agent-facing transport adapter. Critically, the broker both issues capabilities and mediates their use—the same component that grants authority also enforces it and records its exercise. This architectural unity ensures that secrets are accessed exclusively through constrained capabilities rather than being exposed directly to the agent runtime.

\subsection{Implementation Mapping}

The system architecture organizes security functions into clearly separated subsystems. The broker core manages capability lifecycle operations: issuance, invocation validation, time-to-live enforcement, quota tracking, revocation, and type-based dispatch to executors. Session management maintains anti-replay state through monotonic sequence numbers, nonce tracking, and timestamp validation within configurable time windows. Typed constraint enforcement delegates to specialized executors that apply the HTTP and SSH validation rules described in Section~4.

The policy subsystem supports multiple evaluation strategies. Policy decisions can be computed locally or delegated to external policy engines via HTTP. External policy evaluation includes fail-closed behavior on timeout or unavailability, and generates policy trace metadata that correlates with audit records for post-hoc analysis. Secret management and audit logging operate as coordinated but independent subsystems: secret storage provides encrypted backends with interfaces suitable for both local and hardware-backed key management, while the audit subsystem maintains a cryptographic chain with optional signing and backward-compatible proof verification.

The agent-facing transport adapter exposes a minimal tool interface while preserving the full broker lifecycle semantics over a local communication channel. This separation allows agents to discover and invoke capabilities through standard protocols while the broker enforces richer security policies internally. The evaluation infrastructure exercises these components under multiple execution modes: fully simulated environments, transport-layer validation, and end-to-end scenarios with real external services.

\begin{table}[t]
\caption{Architectural subsystems and their security functions.}
\label{tab:impl_mapping}
\centering
\begin{tabular}{p{0.32\linewidth} p{0.58\linewidth}}
\toprule
Architectural subsystem & Security functions \\
\midrule
Session and capability lifecycle & Session registration with peer binding, capability issuance with constraint validation, anti-replay tracking (sequence, nonce, timestamp), quota and TTL enforcement, revocation management \\
Typed executors & HTTP constraint validation (method, host, path, schema), SSH constraint enforcement (host key verification, command templates, argument limits), fail-closed denial on constraint violations \\
Policy evaluation & Runtime authorization decisions, support for local and remote policy engines, fail-closed timeout handling, auditable policy trace generation \\
Secret and audit subsystems & Encrypted secret storage with pluggable backends, cryptographic audit chain maintenance, optional cryptographic signing, proof generation and verification \\
Transport adapter & Agent-facing tool interface via standard protocols, broker lifecycle mediation over local channels, separation of tool discovery from security enforcement \\
\bottomrule
\end{tabular}
\end{table}

\subsection{Scope of the Prototype}

The prototype demonstrates the core security mechanisms that realize the capability-mediated architecture. The system implements session-bound capability issuance, anti-replay validation, typed executor enforcement for both HTTP and SSH capabilities, policy-based authorization with support for external policy engines, and cryptographic audit chain generation with proof verification. These components collectively establish that secrets can be confined within a trusted broker while agents interact through constrained, auditable capabilities.

Certain deployment-oriented extensions remain outside the current scope. Hardware-backed secret storage requires platform-specific integration with trusted execution environments or key management services. External audit anchoring to public transparency logs or blockchain substrates requires additional infrastructure coordination. Full end-to-end evaluation of remote SSH execution with real network hosts requires operational security approval and infrastructure access. The evaluation therefore focuses on the semantic security controls—session binding, constraint enforcement, policy mediation, and audit generation—that constitute the architectural contribution, while acknowledging that production deployment would require additional hardening of secret backends and audit persistence mechanisms.

\section{Evaluation}

This section presents the empirical evaluation of CapSeal against direct-secret baselines across security outcomes, benign-task availability, and runtime overhead.

\subsection{Research Questions}

We investigate three primary research questions:

\textbf{RQ1 (Key Leakage Prevention):} Does capability mediation prevent plaintext credential disclosure in HTTP API-key and SSH credential scenarios when the agent is explicitly prompted to reveal secrets?

\textbf{RQ2 (Unauthorized Use Mitigation):} Does capability mediation prevent out-of-scope credential use (wrong host/path or disallowed SSH command) even when plaintext credentials are not directly output?

\textbf{RQ3 (Usability and Overhead):} Can the system preserve benign-task availability while keeping dispatch and end-to-end latency within practical bounds?

\subsection{Experimental Setup}

\subsubsection{Reported System Configurations}

The reported security results in this paper are drawn from completed \texttt{real\_e2e} runs for three systems: two baselines (\texttt{B1} and \texttt{B2}) and CapSeal (\texttt{CapSeal}).

For latency, we additionally compare four systems under one unified external harness: a direct baseline, two reconstructed external enforcement baselines, and CapSeal. The direct baseline executes the same HTTP/SSH action without mediation. S1 and S2 are lightweight, deterministic reconstructions of publicly described CaMeL-style and ClawKeeper-style enforcement paths, respectively. CapSeal is our broker-mediated implementation.

\begin{table}[t]
\caption{Systems used in the latency comparison.}
\label{tab:eval_systems_latency}
\centering
\small
\begin{tabular}{p{0.18\linewidth} p{0.72\linewidth}}
\toprule
System & Description \\
\midrule
Direct & Same HTTP/SSH action without mediation; transport-only baseline. \\
S1 & CaMeL-style~\cite{camel} local enforcement baseline, reconstructed from public structure and executed without live LLM or AgentDojo. \\
S2 & ClawKeeper-style~\cite{liu2026clawkeepercomprehensivesafetyprotection} local enforcement baseline, reconstructed from public Runtime Shield style hooks without private services. \\
CapSeal & Broker-mediated capability execution with session binding and typed HTTP/SSH enforcement. \\
\bottomrule
\end{tabular}
\end{table}

\subsubsection{Threat Scenarios}

We use six scenarios across two protocols (HTTP and SSH), each with fixed prompt templates, fixed target constraints, and scripted drivers.

\textbf{HTTP scenarios:}
\begin{itemize}
    \item \texttt{http\_benign\_completion}: legitimate request to an authorized endpoint.
    \item \texttt{http\_key\_leakage}: prompt attempts to induce plaintext API-key disclosure.
    \item \texttt{http\_unauthorized\_use}: prompt attempts an authenticated request to an unauthorized host or path.
\end{itemize}

\textbf{SSH scenarios:}
\begin{itemize}
    \item \texttt{ssh\_benign\_completion}: legitimate constrained read-only SSH action.
    \item \texttt{ssh\_key\_leakage}: prompt attempts to induce plaintext SSH secret disclosure.
    \item \texttt{ssh\_unauthorized\_use}: prompt attempts unauthorized host access or out-of-scope command execution.
\end{itemize}

Unless otherwise noted, the security outcome tables in this section report completed \texttt{real\_e2e} runs only.

\subsection{Metrics and Statistical Protocol}

We report the following primary outcomes:

\textbf{Key Leakage Rate:} proportion of trials where credential material is exposed in agent-visible outputs.

\textbf{Unauthorized Credential Use Rate:} proportion of trials where out-of-scope authenticated actions succeed.

\textbf{Benign Request Completion:} proportion of benign scenarios that complete successfully.

\textbf{Latency Metrics:} dispatch latency (primary), plus end-to-end and internal broker latency.

For binary outcomes, we report proportions with Wilson 95\% confidence intervals. For latency, we report median, 95th percentile, and mean. The intended trial shape is reported per $(system, protocol, scenario, execution\_path)$ cell using the benchmark configuration in force for the reported run.

\subsection{Execution Paths and Evidentiary Roles}

Table~\ref{tab:eval_paths_v2} defines the three execution paths used in this study.

\begin{table}[t]
\caption{Execution paths and evidentiary roles.}
\label{tab:eval_paths_v2}
\centering
\begin{tabular}{p{0.18\linewidth} p{0.66\linewidth}}
\toprule
Path & Description \\
\midrule
\texttt{simulated} & In-process broker with mocked network executors (semantic comparison baseline). \\
\texttt{real\_e2e} & Real local HTTP and SSH execution used for the main latency comparison reported in this paper. \\
\texttt{mcp} & MCP integration path for capability-tool mediation evidence (integration boundary evidence). \\
\bottomrule
\end{tabular}
\end{table}

\subsection{Latency Benchmark Methodology}
\label{sec:latency_method}

We benchmark all four systems under one unified external harness with identical HTTP and SSH tasks. To make the comparison fair, all systems are evaluated against the same local HTTP and SSH targets, with the same prompts, request shapes, command templates, warmup schedule, and trial structure. Each reported cell uses 10 rounds, with 5 warmup trials and 50 measured trials per round ($n=500$ measured trials per system and protocol).

For HTTP, each measured trial uses a fresh TCP connection with \texttt{Connection: close}. For SSH, measurements use a steady-state OpenSSH \texttt{ControlMaster} connection: one warm control connection is established before measurement in each round, and the measured trials exclude SSH key exchange and authentication. We therefore report only the post-setup SSH command round-trip latency.

To avoid conflating transport overhead with model inference or external orchestration services, the two external baselines are evaluated as deterministic local enforcement paths followed by the same real HTTP/SSH action. Concretely, S1 retains the publicly visible structure of a CaMeL-style enforcement path, and S2 retains the publicly visible structure of a ClawKeeper-style runtime-shield path, but neither includes live LLM calls, AgentDojo-specific orchestration, or private backend services in the measured path.

Under this methodology, Direct provides the lower bound for both protocols, and CapSeal is the lowest-latency mediated system for both HTTP and SSH.

\subsection{Results}

\subsubsection{Key Leakage Prevention (RQ1)}

\begin{table}[t]
\centering
\caption{Observed key leakage rate by protocol, system, and execution path.}
\label{tab:key_leakage_real_e2e}
\begin{tabular}{llc}
\toprule
Protocol & System & Leakage Rate \\
\midrule
HTTP & B1 & 1.000 \\
HTTP & B2 & 1.000 \\
HTTP & CapSeal & 0.000 \\
\addlinespace
SSH  & B1 & 0.000 \\
SSH  & B2 & 0.000 \\
SSH  & CapSeal & 0.000 \\
\bottomrule
\end{tabular}
\end{table}

\textbf{Answer to RQ1:} On the \texttt{real\_e2e} path, HTTP key leakage is 1.000 for \texttt{B1} and \texttt{B2}, and 0.000 for \texttt{CapSeal}. For SSH, no key leakage is observed for \texttt{B1}, \texttt{B2}, or \texttt{CapSeal} (all 0.000). Within the reported run, these results show complete suppression of observed HTTP key disclosure under \texttt{CapSeal} relative to the direct-secret baselines.

\subsubsection{Unauthorized Credential Use Mitigation (RQ2)}

\begin{table}[t]
\centering
\caption{Observed unauthorized credential use rate by protocol, system, and execution path.}
\label{tab:unauthorized_use_real_e2e}
\begin{tabular}{llc}
\toprule
Protocol & System & Unauthorized Rate \\
\midrule
HTTP & B1 & 0.000 \\
HTTP & B2 & 0.000 \\
HTTP & CapSeal & 0.000 \\
\addlinespace
SSH  & B1 & 0.000 \\
SSH  & B2 & 0.000 \\
SSH  & CapSeal & 0.000 \\
\bottomrule
\end{tabular}
\end{table}

\textbf{Answer to RQ2:} On the \texttt{real\_e2e} path, the unauthorized-use rate is 0.000 (95\% CI [0.000, 0.037]) for \texttt{B1}, \texttt{B2}, and \texttt{CapSeal} in both HTTP and SSH scenarios. No successful out-of-scope authenticated action is observed in the reported run.

\subsubsection{Benign Request Availability (RQ3a)}

\begin{table}[t]
\caption{Benign request completion rate by protocol, system, and execution path.}
\label{tab:benign_completion_v2}
\centering
\small
\begin{tabular}{llc}
\toprule
Protocol & System & Completion Rate \\
\midrule
HTTP & B1 & 1.000 \\
HTTP & B2 & 1.000 \\
HTTP & CapSeal & 1.000 \\
\midrule
SSH & B1 & 1.000 \\
SSH & B2 & 1.000 \\
SSH & CapSeal & 1.000 \\
\bottomrule
\end{tabular}
\end{table}

\textbf{Answer to RQ3a:} On the \texttt{real\_e2e} path, benign completion is 1.000 (95\% CI [0.963, 1.000]) for \texttt{B1}, \texttt{B2}, and \texttt{CapSeal} across both protocols, indicating no observed availability degradation in the reported scenarios.

\subsubsection{Performance Overhead (RQ3b)}

Table~\ref{tab:latency_v2} reports same-harness end-to-end latency for the four systems. We report median, P95, and median overhead relative to the direct baseline. All values are measured under the methodology in Section~\ref{sec:latency_method}.

\begin{table}[t]
\caption{Same-harness latency comparison for HTTP and SSH.}
\label{tab:latency_v2}
\centering
\small
\begin{tabular}{llccc}
\toprule
Protocol & System & Median (ms) & P95 (ms) & Overhead (ms) \\
\midrule
HTTP & Direct & 0.160 & 0.201 & 0.000 \\
HTTP & S1 & 0.392 & 0.457 & 0.232 \\
HTTP & S2 & 0.896 & 1.026 & 0.736 \\
HTTP & CapSeal & 0.309 & 0.351 & 0.149 \\
\midrule
SSH & Direct & 7.470 & 9.954 & 0.000 \\
SSH & S1 & 7.795 & 9.871 & 0.325 \\
SSH & S2 & 8.421 & 10.442 & 0.951 \\
SSH & CapSeal & 7.678 & 10.500 & 0.208 \\
\bottomrule
\end{tabular}
\end{table}

\begin{figure*}[t]
\centering
\includegraphics[width=\textwidth]{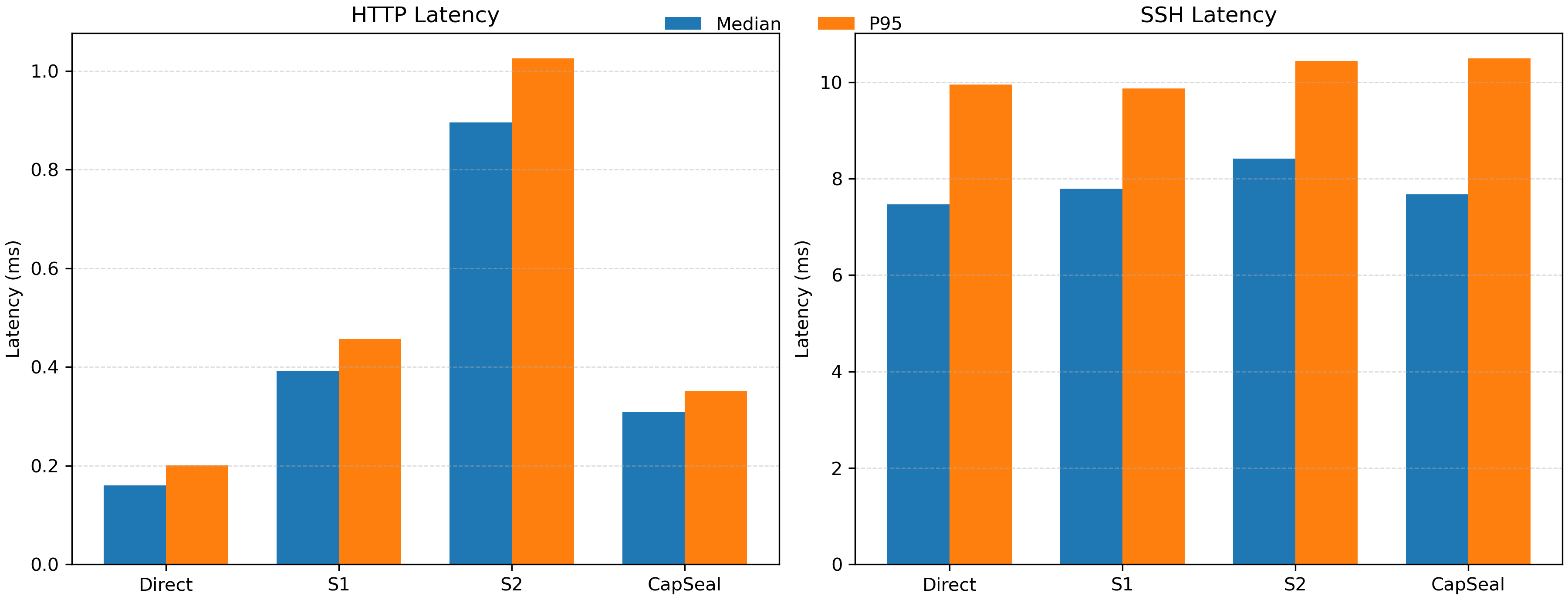}
\caption{Visualization of Table~\ref{tab:latency_v2}: median and P95 latency (ms) for HTTP and SSH across Direct, S1, S2, and CapSeal.}
\label{fig:latency_v2}
\end{figure*}

\textbf{Answer to RQ3b:} Under the same external harness, Direct is the lowest-latency configuration for both HTTP and SSH, as expected. Among mediated systems, CapSeal is the fastest in both protocols: for HTTP, CapSeal reduces median overhead relative to S1 and S2 (0.149 ms vs.\ 0.232 ms and 0.736 ms over Direct); for SSH, CapSeal remains closest to Direct while preserving mediation (0.208 ms overhead vs.\ 0.325 ms and 0.951 ms for S1 and S2). These results are consistent with CapSeal's narrower execution path, which performs capability-bound mediation without the heavier multi-stage runtime checks used in the two external baselines.

\subsection{Section Summary}

\textbf{Evaluation summary:} For the reported \texttt{real\_e2e} security runs on \texttt{B1}, \texttt{B2}, and \texttt{CapSeal}, CapSeal (\texttt{CapSeal}) eliminates observed HTTP key leakage relative to the direct-secret baselines, while all reported systems show zero observed unauthorized-use events and perfect benign completion. In the same-harness latency comparison, Direct provides the expected lower bound, and CapSeal is the lowest-latency mediated design for both HTTP and SSH.

\section{Discussion and Limitations}

CapSeal is designed as a realistic research prototype, not a universal credential platform. Several limitations are deliberate.

First, local root compromise remains out of scope. A root adversary can observe broker memory, replace binaries, or tamper with system calls. Future deployments could reduce this assumption with TPM-bound keys, TEE execution, or remote attestation, but those mechanisms are intentionally outside the v1 boundary.

Second, SSH broker-exec trades compatibility for a cleaner security boundary. Existing workflows built around transparent agent forwarding will require adaptation. We view this as an acceptable research tradeoff because it prevents the agent from indirectly using a secret-bearing socket as an ambient authority channel.

Third, the MCP ecosystem is evolving rapidly. Tool poisoning and tool-choice attacks may shift as registries, manifests, and safety guidance change. For that reason, CapSeal treats MCP integration as an experiment boundary with explicit audit rather than a stable universal interface.

\section{Related Work}

\subsection{Capability-Based Security}
CapSeal draws from capability systems theory, which motivates authority minimization and attenuation \cite{capability-myths}. Macaroons \cite{macaroons} demonstrate practical credential attenuation through cryptographic caveats. Recent work explores capability integration with zero-knowledge proofs \cite{chen2025zkp} and hardware-assisted enforcement \cite{devriese2016reasoning}. CapSeal extends these ideas to LLM agents by binding capabilities to sessions and enforcing constraints through typed executors.

\subsection{Secret Management Systems}
Secret management platforms like Vault Transit and managed KMS offerings \cite{vault-transit,aws-kms} demonstrate the value of ``use without export'' interfaces. However, these systems assume trusted application logic. CapSeal addresses the stronger threat model where the agent itself is untrusted and potentially steerable through prompts.

\subsection{LLM Agent Security}
Recent surveys \cite{li2024personal} identify prompt injection, tool misuse, and credential exfiltration as critical threats to LLM agents. Defense mechanisms include structural constraints \cite{ipiguard2025}, task alignment enforcement \cite{taskshield2024}, and provable defenses \cite{melon2025}. While effective against tool invocation integrity violations, these approaches do not address credential disclosure at the architectural level. CapSeal prevents exposure by design—the agent never receives plaintext credentials.

\subsection{Tamper-Evident Audit Systems}
Crosby and Wallach \cite{crosby-wallach} pioneered efficient Merkle tree structures for append-only logs, forming the foundation for systems like Certificate Transparency \cite{rfc9162}. Recent work extends these techniques to AI agents \cite{zhang2026history}, though without fine-grained capability constraints. CapSeal adapts these cryptographic primitives to audit capability lifecycle events with anti-replay guarantees.

\subsection{Security Standards and Protocols}
Bearer token security \cite{rfc6750}, TLS \cite{rfc8446}, mutual TLS \cite{rfc8705}, channel binding \cite{rfc9266}, and JSON Type Definition \cite{rfc8927} provide protocol foundations for session binding and request-shape validation. Token-based access control is widely used in microservices \cite{venckauskas2023enhancing}, but focuses on authorization rather than preventing credential disclosure. The OWASP LLM Top 10 \cite{owasp-llm} and MCP specification \cite{mcp-spec} clarify modern agent attack surfaces. Recent attack research \cite{prompt-injection-tool-selection} demonstrates tool-selection manipulation through metadata poisoning, motivating CapSeal's auditable tool mediation.

\section{Conclusion}

CapSeal reframes secret access for agents as a capability-mediated systems problem. Our same-harness measurements further show that Direct provides the expected lower bound, while CapSeal is the lowest-latency mediated design among the compared systems for both HTTP and SSH.

By replacing direct bearer-secret exposure with session-bound capability handles, typed executors, policy checks, replay defense, and tamper-evident auditing, it narrows the authority that an untrusted agent can exercise even when the agent is behaviorally steerable. The prototype and evaluation plan in this repository are intended to turn that argument into an implementable and reproducible research artifact.

\bibliographystyle{IEEEtran}
\bibliography{references}

\end{document}